# Integrating quantum and classical computing for multi-energy system optimization using Benders decomposition

Ludger Leenders[a,1], Martin Sollich [a,b,1], Christiane Reinert [b], André Bardow[a,*]

[a] Energy & Process Systems Engineering, Department of Mechanical and Process Engineering, ETH Zürich, Tannenstrasse 3, 8092 Zürich, Switzerland

[b] Institute of Technical Thermodynamics, RWTH Aachen University, Schinkelstraße 8, 52062 Aachen, Germany

**Abstract:** During recent years, quantum computers have received increasing attention, primarily due to their ability to significantly increase computational performance for specific problems. Computational performance could be improved for mathematical optimization by quantum annealers. This special type of quantum computer can solve quadratic unconstrained binary optimization problems. However, multi-energy systems optimization commonly involves integer and continuous decision variables. Due to their mixed-integer problem structure, quantum annealers cannot be directly used for multi-energy system optimization.

To solve multi-energy system optimization problems, we present a hybrid Benders decomposition approach combining optimization on quantum and classical computers. In our approach, the quantum computer solves the master problem, which involves only the integer variables from the original energy system optimization problem. The subproblem includes the continuous variables and is solved by a classical computer. For better performance, we apply improvement techniques to the Benders decomposition. We test the approach on a case study to design a cost-optimal multi-energy system. While we provide a proof of concept that our Benders decomposition approach is applicable for the design of multi-energy systems, the computational time is still higher than for approaches using classical computers only. We therefore estimate the potential improvement of our approach to be expected for larger and fault-tolerant quantum computers.

**Keywords**:
*Quantum Annealing, MILP, Sector Coupling, Hybrid Algorithm*

**Highlights:**

- Quantum computing for multi-energy system optimization
- Benders decomposition using quantum computers
- Hybrid optimization combining quantum and classical computer
- Master problem solved on quantum computer
- Quantum approach feasible but outperformed by classical computers

[1] These authors contributed equally.

[*] Corresponding author: A. Bardow, Energy & Process Systems Engineering, Department of Mechanical and Process Engineering, ETH Zürich, Tannenstrasse 3, 8092 Zürich, Switzerland
E-mail: abardow@ethz.ch

# 1 Introduction

Multi-energy systems allow for sector-coupling in industry, which is crucial for decarbonization. However, sector-coupling also renders multi-energy system optimization highly complex. Furthermore, wind and solar power result in high spatiotemporal fluctuations (Ridha et al., 2020). Due to these challenges, operation and design of multi-energy systems is best addressed by mathematical optimization (Andiappan, 2017). Most commonly, optimization problems of multi-energy systems are written as Mixed-Integer Linear Programming (MILP) problems to balance the need for physical accuracy with computational costs (Wirtz et al., 2021).

The computational effort in solving multi-energy system optimization problems increases rapidly with the problem size. Goderbauer et al. (2019) proved that the operational problem of decentralized multi-energy systems is weakly NP-hard while the synthesis that combines operation and design systems is even strongly NP-hard. Thus, the size of energy system optimization problems that can be solved is currently limited and the modeler has to tackle a trade-off between the model complexity and computational efficiency (Kotzur et al., 2021). Hence, solving large instances of multi-energy system optimization remains challenging, since on the hardware side further improvements of classical computers are limited due to the saturation limits of Moore's law, thereby limiting computational tractability (Mack, 2011).

Recently, quantum computers have received increasing attention across research fields. Quantum computers promise significant speed-up for certain computational problems (Grover, 1997; Shor, 1999). The most promising fields are the simulation of quantum systems, cryptography, search, solving large systems of linear equations, and optimization (Montanaro, 2016). However, many promising algorithms require noise-free quantum computers, whereas today's quantum computers are noisy. Researchers say that we are currently in the noisy intermediate-scale quantum (NISQ) era (Brooks, 2019). To work with the available noisy quantum computers, specific algorithms are required.

Today, two main approaches exist for building quantum computers: universal quantum computers and quantum annealers. Both types of quantum computers can solve combinatorical optimization problems. Compared to quantum annealers, universal quantum computers are widely applicable (Arute et al., 2019; Ball, 2021; Zhu et al., 2022). Today's noisy universal quantum computers mainly employ quantum approximate optimization algorithms (QAOA) (Farhi et al., 2014). Current universal quantum computers only have a few hundred qubits (Gooding, 2021; Madsen et al., 2022; Editorial, 2023). The number of qubits relates to the maximal number of binary variables that the problem can incorporate. Thus, universal quantum computers are still very limited in tractable problem size.

In comparison to universal quantum computers, today, the number of qubits is significantly higher for quantum annealers with more than 5000 qubits (McGeoch and Farre, 2020). Quantum annealing can solve combinatorial optimization problems (Hauke et al., 2020; Bernal et al., 2022). In particular, quantum annealers can solve *polynomial* unconstrained binary optimization problems in principle. However, at the current state, quantum annealers can only solve *quadratic* unconstrained binary optimization (QUBO) problems (Hauke et al., 2020). Thus, for solving optimization problems with other than binary variables, they need to be represented using binary variables. Integer variables can be modeled by binary encoding (Chang et al., 2020; Zhao et al., 2022). Continuous variables can also be expressed by binary encoding, however, they have to be discretized (Zhao et al., 2022). The more binary variables are used for discretization, the higher the accuracy. A trade-off occurs between the accuracy of the continuous variable representation and the number of used binary variables, i.e., qubits.

For solving MILPs with quantum computers, one approach discretizes the continuous variables and employs afterwards binary encoding for the integer variables. The approach was successfully applied by Ajagekar and You (2019) for solving a heat exchanger network synthesis problem on a quantum annealer. Furthermore, Halffmann et al. (2023) and Braun et al. (2023) both solve an unit commitment problem on a quantum annealer by reforumalting the original problem into a QUBO. Deng et al. (2023) reformulate a model predictive control problem of a building HVAC into a QUBO.

We conclude that while quantum computers are promising for binary optimization problems, the necessary discretization of continuous variables severely limits their applicability. Since continuous variables are not a concern for classical computers, this challenge can be overcome by coupling classical and quantum computers to solve problems with discrete and continuous variables. Thereby, the advantages of both computers can be retained by combining quantum computers deciding on integer variables with classical computers deciding on continuous variables.

In the literature, combinations of classical and quantum computers are already available for solving Mixed-Integer Programming (MIP) problems. The following approaches use Benders decomposition (Benders, 1962) for this purpose: Zhao et al. (2022) use a quantum annealer to solve the master problem within the Bender's decomposition of an MILP. The authors use a quantum annealer to solve the master problem. Fan and Han (2022) solve a MILP via Bender's decomposition and use a quantum annealer. The solution method is based on Zhao et al. (2022). Chang et al. (2020) solve the integer problems in the Benders decomposition of an MIP by a quantum annealer. In their case study, a unit commitment problem is solved for a power system. Paterakis (2023) uses the quantum annealer for a cut selection procedure within the Bender's decomposition. The algorithm is applied to a unit commitment problem. Gao et al. (2022) solve a MIP unit commitment problem via Bender's decomposition and use a quantum annealer. The reviewed methods were only used to optimize power systems considering electricity as the sole energy form. As stated previously, sector coupling becomes increasingly important. However, approaches are currently missing to optimize more complex multi-energy system designs via quantum and classical computers.

Here, we propose a Benders decomposition to optimally design and operate multi-energy systems using a quantum annealer to optimize integer decisions and a classical computer for continuous decisions. We can summarize the contribution of this work as:

- Benders decomposition for MILP optimization using a quantum annealer including improvement techniques for the Benders decomposition to ensure computational tractability.

- Application of quantum computers for the design optimization of multi-energy systems considering renewable energy conversion technologies and time-coupling constraints via energy storage units.

- Comparison with a state-of-the-art solver and Benders decomposition using classical computers only.

- Estimation of the potential improvements once qunatum computers are larger and fault-tolerant.

The remainder of the paper is organized as follows: Section 2 presents the Benders decomposition, using the quantum annealer and the improvement techniques that we applied for the Benders decomposition. In Section 3, our approach is applied to a case study to design a multi-energy system and compare our method to classical computers. In the final section, we draw our conclusions.

## 2 Benders decomposition for quantum-computing-based optimization

In this section, the approach for Benders decomposition is presented in general, which is partly based on the work of Chang et al. (2020) and Zhao et al. (2022). Firstly, we briefly summarize the Benders decomposition and how to decompose the MILP into a subproblem and a master problem (Section 2.1). Secondly, we reformulate the master problem into a QUBO such that the master problem can be solved by a quantum computer. Finally, improvement techniques are introduced to the Benders decomposition.

### 2.1 Benders decomposition for solving MILPs

We use Benders decomposition to decompose an MILP problem such that the continuous decision variables are only considered in the subproblem and the integer decisions are considered in the master problem. The subproblem is then solved with a classical computer and the master problem using a quantum computer.

In an iterative approach, the subproblem is solved for the given solutions of the master problem. If the subproblem is feasible, we provide feasible solutions of the original MILP. The feasible solutions of the subproblem are used as an upper bound of the original MILP. In our Benders decomposition, the master problem provides lower bounds to the original MILP problem by optimizing the integer variables, while the subproblem is represented by the continuous surrogate variable $\zeta$. In the master problem, the surrogate variable $\zeta$ provides a lower bound on the subproblem objective, which depends on the continuous variables.

The Benders decomposition iterates between master problem and subproblem, until a predefined optimality gap between lower and upper bound is reached. In each iteration, optimality and/or feasibility cuts are added to the master problem. Optimality cuts are generated from feasible subproblems, providing a lower bound on the part of the objective function depending on the continuous variables. Feasibility cuts are generated from infeasible subproblems, thereby excluding the related integer solutions from the master problem.

We can write the original MILP in the general form as follows:

$$\min_{\mathbf{x},\mathbf{y}} \mathbf{c}^T\mathbf{x} + \mathbf{d}^T\mathbf{y} \tag{1}$$

$$\text{s.t. } A\mathbf{x} + B\mathbf{y} \geq \mathbf{b} \tag{2}$$

$$\mathbf{x} \geq 0 \tag{3}$$

$$\mathbf{y} \geq 0 \tag{4}$$

$$\mathbf{x} \in \mathbb{R}^n, \mathbf{y} \in \mathbb{Z}^m \tag{5}$$

$\mathbf{x}$ and $\mathbf{y}$ are the continuous and the integer decision variables, respectively, which are defined to be positive (Eq. 3 and 4). $\mathbf{c}$ and $\mathbf{d}$ are the corresponding cost factors in the objective function. $A$, $B$ and $\mathbf{b}$ are the parameters for the constraints (Eq. 2).

We decompose the original problem (Eq. 1 - 5) into a master problem and a subproblem (Benders, 1962). In the subproblem, we only consider the continuous variables $\mathbf{x}$ while the integer variables $\overline{\mathbf{y}}$ are fixed. Thus, the subproblem is an LP problem of the following form:

$$\min_{\mathbf{x}} \mathbf{c}^T\mathbf{x} + \mathbf{d}^T\overline{\mathbf{y}} \tag{6}$$

$$\text{s.t. } A\mathbf{x} \geq \mathbf{b} - B\overline{\mathbf{y}} \tag{7}$$

$$\mathbf{x} \geq 0 \tag{8}$$

$$\mathbf{x} \in \mathbb{R}^n \tag{9}$$

In the subproblem, we have the same constraints (Eq. 7-9) as in the original MILP, while the integer variables $\overline{\mathbf{y}}$ are fixed.

In Benders decomposition, the dual of the subproblem is generated and then solved. The dual subproblem is:

$$\max_{\mathbf{v}} (\mathbf{b} - B\overline{\mathbf{y}})^T\mathbf{v} + \mathbf{d}^T\overline{\mathbf{y}} \tag{10}$$

$$\text{s.t. } A^T\mathbf{v} \leq \mathbf{c} \tag{11}$$

$$\mathbf{v} \geq 0 \tag{12}$$

$$\mathbf{v} \in \mathbb{R}^o \tag{13}$$

Thereby, $\mathbf{v}$ are the dual variables associated with the constraints in Eq. (7).

The master problem optimizes the integer variables. The surrogate variable $\zeta$ acts as a surrogate for the dual subproblem in the master problem:

$$\min_{\mathbf{y},\zeta} \zeta + \mathbf{d}^T\mathbf{y} \tag{14}$$

$$\text{s.t. } \overline{B}\mathbf{y} \geq \overline{\mathbf{b}} \tag{15}$$

$$(\mathbf{b} - B\mathbf{y})^T v_o \leq \zeta \quad \forall o \in O \tag{16}$$

$$(\mathbf{b} - B\mathbf{y})^T u_f \leq 0 \quad \forall f \in F \tag{17}$$

$$\mathbf{y} \geq 0 \tag{18}$$

$$\zeta \in \mathbb{R}, \mathbf{y} \in \mathbb{Z}^m \tag{19}$$

In the master problem, we consider all constraints that are pure integer (Eq. 15). Eq. (16) comprises the optimality cuts. Optimality cuts are generated from the optimal solutions of the dual subproblem $v_o$. The optimality cuts provide a lower bound to the surrogate of the subproblem $\zeta$. Thus, the objective function (Eq. 14) of the master problem is a lower bound of the original MILP problem.

Since the Benders decomposition starts with solving the master problem, the surrogate $\zeta$ is unbounded in the first iteration of the Benders decomposition. We generate a first optimality cut by solving the LP relaxation of the original MILP problem to bound $\zeta$ in the first iteration.

During the iterations, the master problem can provide a solution that is infeasible in the original MILP problem. Such a solution yields an infeasible subproblem. The infeasibility implies an unbounded dual subproblem based on the duality theory. To exclude infeasible solutions in the next iteration of the master problem feasibility cuts are generated (Eq. 17). The vector $u_f$ represents a direction of unboundedness of the dual problem for the current integer solution.

Our Benders decomposition approach solves the master problem on a quantum annealer. Since quantum annealer can currently only solve QUBOs, we need to reformulate the master problem. Thus, we need to discretize $\zeta$ and reformulate the problem such that all constraints become a part of the objective function. We briefly present the reformulation in the next section.

## 2.2 Master problem reformulation for quantum computers

First, the reformulation of the surrogate variable $\zeta$ using binary variables is described. Second, we reformulate the master problem into an unconstrained optimization problem, yielding the final QUBO formulation.

### 2.2.1 Reformulation of surrogate variable $\zeta$

All decision variables from the original master problem are integer, except the surrogate of the subproblem objective $\zeta$, which is continuous. Following (Chang et al., 2020; Zhao et al., 2022), we discretize $\zeta$ by binary encoding to model $\zeta$ by using only binary variables $f_k$:

$$\zeta \cong \sum_{k=0}^{k^{max}} d \cdot 2^k \cdot f_k \tag{20}$$

For the binary encoding, we estimate the maximal value of the variable beforehand to determine $k^{max}$. We want to keep the value range of the variable as small as possible to keep the number of binary variables low, due to limited size of the quantum annealer. For the surrogate variable $\zeta$, this can be done by naively generating non-optimal solutions in advance and reading the continuous part of the objective function value. If the parameter $d$ is 1, the reformulation is used for integer variables. For decimal numbers like the surrogate variable $\zeta$, the discretization rate $d$ is adapted, e.g., 0.01 for two decimals.

To represent integer variables as binary variables, we also use binary encoding. The resulting master problem contains only binary variables but is still not a quadratic unconstrained binary optimization (QUBO) problem, the problem type needed by today's quantum annealers.

### 2.2.2 Transformation of the master problem into a QUBO

The transformation of an optimization problem involving only binary variables into a QUBO is well known and standard practice for using quantum computers (Glover et al., 2018). Thus, we only briefly describe the procedure.

For the transformation, a penalty term is added to the objective function for each constraint, such that the violation of the constraint is penalized. For this purpose, all equality constraints are reorganized such that we have all terms on the left-hand side and the right-hand side is 0. Then, the left-hand side is squared and added to the objective function. Thereby, the added term is always positive or 0. As a consequence, the optimizer prefers to decrease the term in a minimization problem and wants to fulfill the constraint by reaching 0.

We first have to transform all inequality constraints into equality constraints by introducing slack variables. The slack variables are also modeled by binary encoding similarly to the surrogate variable $\zeta$. Afterwards, these constraints are added to the objective function as well via the described penalization approach.

For each constraint its penalty term in the objective function is scaled by a penalty coefficient. The penalty coefficient adjusts the influence of the constraint violation. For example, if the solution of a QUBO violates a certain constraint, a higher penalty coefficient is chosen for this constraint to reduce its violation. However, if a penalty coefficient is increased, the importance of the original objective function is decreased. Thus, suitable penalty coefficients must be identified for each problem at hand (Glover et al., 2022).

## 2.3 Improvement techniques applied in Benders decomposition

In addition to the classical Benders decomposition (Section 2.1), we apply several techniques to improve the computational performance. We apply the multi-cut approach, valid inequalities, LP relaxation of the master problem, and LP relaxation of the original MILP, which we describe in further detail below.

### Multi-cut approach

The multi-cut approach adds not one but several optimality or feasibility cuts to the master problem in each iteration. For this purpose, more than one solution of the master problem is generated, which can be easily done by modern classical solvers, such as Gurobi, as well as by the quantum annealer from D-Wave. For each solution, the related subproblem is solved and either an optimality or a feasibility cut is generated. The multi-cut approach typically reduces the number of iterations needed while increasing the time needed per iteration. Whether the multi-cut approach decreases the overall time for the Benders decomposition, depends on the problem itself and cannot be generalized (You and Grossmann, 2013). Especially when using a quantum computer with very limited size as today's quantum computer, the increase of the size of the QUBO can even cause worse behavior. For our multi-energy system optimization in the case study, we assessed for each problem instance separately if the multi-cut approach was beneficial. It was beneficial for some problem instances but not for all of them. If it was beneficial, two or three cuts per iteration provided the best improvement.

### Valid inequalities

Valid inequalities are constraints that are not necessary for the problem definition, but provide additional bounds on the solution space of the master problem. These additional constraints could also be identified by the Benders decomposition as feasibility or optimality cuts but identifying these constraints along the iterations could cause long convergence times. Therefore, including valid inequalities can improve convergence (Cordeau et al., 2000).

An example of valid inequalities used in this study is that the summed capacity of the installed energy supply units is at least large enough to supply the peak demand. For our case study (Section 3), the use of the valid inequality was not beneficial in terms of convergence behavior, if the formulation of a valid inequality required the definition of additional variables.

### LP relaxation of the master problem

Solving the LP relaxation of the master problem is less computationally expensive than solving the master problem itself. From the solution of the relaxed master problem, valid cuts can be generated. In the best case, these cuts could be more beneficial than the cuts obtained by solving the master problem. However, whether these cuts from the relaxed master problem improve the convergence behavior depends again on the problem instance at hand (McDaniel and Devine, 1977). Based on preliminary tests, we apply this improvement technique for the first iteration of our Benders decomposition.

### LP relaxation of the original MILP

We solve the LP relaxation of the original MILP to generate initial optimality cuts for the master problem. Thereby, we avoid defining an arbitrary lower bound of the surrogate variable $\zeta$ for the first iteration of the Benders decomposition. Additionally, if the lower bounds provided by the master problem are worse than the solution of the LP relaxation of the original MILP, the LP relaxation of the original MILP can provide tighter lower bounds. Also, in case the master problem is solved with a heuristic or as in our case study with an error-prone quantum computer, the solution from the master problem cannot be used as a lower bound since it requires guaranteed global optimality.

## 3 Case Study: Multi-energy system design optimization

In our case study, the presented Benders decomposition is used to design a multi-energy system. Compared to power system problems, multi-energy system problems often have a sparser constraint matrix, since multi energy forms are modeled. This can influence the performance of the optimization problem. The multi-energy system design problem is based on Baumgärtner et al. (2019) and implemented in the SecMOD framework from Reinert et al. (2023, 2022).

The energy system has to supply a temporally resolved heating, electricity, and cooling demand. Potential components are: Heat pump, compression chiller, combined-heat-and-power engine, photovoltaic, and battery (c.f. Figure 1). The design of the energy system is optimized while considering the optimal operation. The batteries introduce time-coupling in the operational optimization of the multi-energy system. The binary decision variables describe if a unit is purchased, if the storage charges or discharges, and part-load curves of the components. The TSAM tool is used for time-series aggregation (Kotzur et al., 2018) to aggregate the hourly energy demands to 2-5 typical time steps (Table 1). We solve the design problem until an optimality gap of 5 % is reached. We use such a high tolerance since with today's quantum annealers it is not guaranteed that the global optimal solution is found. In fact, even if the global optimal solution is found, it cannot be identified as the global optimum, since a static lower bound must be used to calculate the optimality gap as the error-prone quantum annealer does not allow the retrieval of lower bounds from solving the master problem. This lower bound is obtained by solving the LP-relaxation of the original MILP. For the discretization of the continuous variables in the master problem, we choose an accuracy of $10^{-1}$. In comparison, the objective function is within the magnitude of $10^5$. We assessed the discretization before using the quantum computer via the Benders-SimA optimization approach, since we only had limited access to the quantum computer. We did not observe any gains by increasing the accuracy beyond $10^{-1}$. For the same reason, we also assessed and chose the penalty parameters using the Benders-SimA before using the quantum computer.

The data handling, the call of the optimization solvers and quantum annelaer is written in Python 3.8. The subproblems are solved with Gurobi 9.5.1 (Gurobi Optimization, 2022) on an Intel Core i7. The master problem is solved with the D-Wave - Advantage system 4.1 on AWS, which has 5,627 qubits and 40,279 couplers. To solve the QUBO at hand with the quantum annealer, the QUBO must be mapped on the given qubit graph structure. This process is called embedding. Here, we use the minorminer embedding function provided by D-Wave. The qubits within the Advantage system are not fully connected, which increases the required number of qubits the QUBO at hand must be mapped on. Due to the limited size of the quantum annealer as well as the limited interconnection of the qubits, we set the solver limit of the maximal size of the QUBOs solved on the quantum annealer to 150 to 170 binary variables, depending on the instance. If a QUBO is larger, the QBSolv solver provided by D-Wave decomposes the QUBO into smaller sub-QUBOs which can then be solved by the quantum annealer.

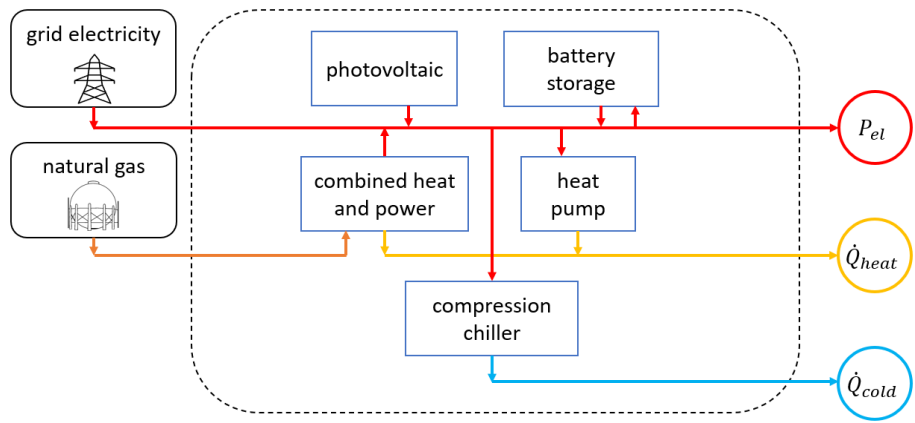

**Fig. 1.** Superstructure of the energy system designed in the case study.

**Tab. 1.** Problem instances of the case study and their distinct characteristics.

| time steps | constraints | variables | continuous variables | binary variables |
|---|---|---|---|---|
| 2 | 154 | 41 | 21 | 20 |
| 3 | 230 | 59 | 31 | 28 |
| 4 | 294 | 77 | 41 | 36 |
| 5 | 374 | 95 | 51 | 44 |

As Ding et al. (2021) pointed out, the computing time on the quantum annealer is almost negligible, but solving a QUBO with the quantum annealer can still be very time-consuming due to the queuing time. Since the queuing time depends on the current traffic on the cloud and the demand for D-Wave's quantum annealers, those times are not consistent and, therefore, difficult to interpret. Hence, we focus on the actual quantum computation time as the solution time for the master problem. The actual quantum computing time is provided by D-Wave via the attribute QPU access time.

Since quantum annealers are still noisy, the annealing procedure is executed multiple times for the same problem. Based on a small parameter sensitivity analysis, the number of reads, i.e., annealing procedures executed on the quantum annealer, is set to values between 50 and 300 in our case study. The number of repeats defines how often the QBSolv algorithm is repeated with no change in the optimal value. In our case study, we used values between 5 and 20.

To compare the performance of our approach, we benchmark our approach (Benders - QA) by:

- Solving the original MILP without decomposition using Gurobi 9.5.1 (Gurobi)

- Using Benders decomposition approach and solving master problem and subproblem using Gurobi 9.5.1 (Benders-Gurobi)

- Using Benders decomposition approach applying the simulated annealing heuristic from D-Wave on a classical computer for the master problem (D-Wave, 2022) and Gurobi 9.5.1 for the subproblem (Benders–SimA).

We chose these benchmarks, since Gurobi is a state-of-the-art commercial solver for MILP problems. However, Bender decomposition can be more efficient in solving very large problem instances, in comparison to solving the MILP directly via Gurobi, although we did not solve such large instances in this work. To assess the impact of the quantum annealer on the Benders decomposition, we compare it to Benders-Gurobi. Furthermore, by comparing to Benders decomposition with simulated annealing, we compare to another annealing approach.

## 3.1 Computational Results

Our Benders decomposition approach using the quantum annealer identifies an optimal solution within the predefined optimality gap of 5 %. The multi-energy systems requires a heat pump, a compression chiller and a combined-heat-and-power unit in all considered cases (Table 2). The photovoltaic unit is sized similar in all cases, whereas the battery is slightly differently, but always sized small. We also compare the objective function values and observe that our approach identifies a near-optimal solution within the optimality gap (Table 2). Thus, our approach identifies similar multi-energy systems as the original MILP solved via Gurobi. However, the Benders decomposition using the quantum annealer has the highest computation time in all cases (Figure 2). Using the MILP solver Gurobi to directly solve the original MILP has the lowest computation time for all cases. The second fastest computation time is always reached with the Benders - Gurobi approach, followed by the Benders - SimA approach. The computation times of the different approaches increase with problem size, given here by the number of typical time steps. Only for our Benders decomposition, the computation time decreases between 3 and

**Tab. 2.** Multi-energy system design for different number of typical time steps. The solutions are presented by the Benders QA approach and from solving the original MILP with Gurobi (in brackets). The symbol "-" indicates that the component was not built. The optimality gap is calculated by $(best\,found\,solution - Gurobi\,solution)/Gurobi\,solution$. The *best found solution* refers to the solution obtained by the Bender's - QA approach, respectively for each instance.

| time steps | Heat pump (MW) | Compression chiller (MW) | Combined heat and power (MW) | Photovoltaic unit (MW) | Battery (MWh) | Optimality gap (%) |
|---|---|---|---|---|---|---|
| 2 | 1 (1) | 4.5 (4.5) | 3.2 (3.2) | - (-)   | - (-)   | 0   |
| 3 | 1 (1) | 5 (4.4)   | 3.2 (3.1) | 10 (10) | - (0.3) | 1.8 |
| 4 | 1 (1) | 6 (6)     | 3.2 (3.2) | 0.1 (0) | 0.1 (0) | 0.8 |
| 5 | 1 (1) | 5 (4.3)   | 3.1 (3.2) | 10 (10) | - (0.2) | 2.2 |

4 typical time steps. This decrease is caused by the fact that we check the optimality gap after each repeat on the quantum annealer. As already mentioned, we run the QBSolv algorithm multiple times. If the gap is sufficiently closed after an early run, our algorithm stops. This stopping occurs early when we optimize 4 typical time steps. For the instance with 2 time steps, 6 out of the 154 original constraints were pure integer constraints and were therefore added to the master problem, while the others were added to the subproblem. For the same instance, 23 additional pure integer constraints were added to the master problem as valid inequalities.

We identify multiple reasons for the high computation times in the Benders quantum annealer approach (Figure 3): First, the used quantum annealer has a solver limit for a maximum number of binaries per QUBO, which implies that the computation time increases when the QUBO needs to be decomposed. In our case, the QUBO of the master problem always needs to be decomposed. Even for the case with only 2 typical time steps, where the master problem has only 21 binary variables, we still have to reformulate the surrogate variable $\zeta$ and the slack variables introduced by the inequality constraints, which increases the total number of binary variables in the master problem. Second, the data processing between the different optimization problems consumes much computation time (Figure 3). Here, substantial improvements are possible by more efficient data handling. We use Python as a programming language and switching to another language e.g., C++, the code would most likely execute much faster as shown by Fua and Lis (2020). They showed python running times that were more than a factor of 10 larger than the ones of C++, while the gap further increased for increasing problem sizes. The third reason is that today's quantum annealers are not flawless and have a non-deterministic solution behavior. Thus, the number of repeats is set to 5-20 times depending on the problem instance. As a consequence, the solution time of the master problem could be reduced by this factor, when the hardware of quantum annealers would be flawless.

We can estimate a potential best case computational time for our approach by assuming fault-tolerant quantum annealers with at least as many qubits available such that the master problem can be mapped entirely. We derive the time of a single quantum task from our previous experiments and assume that a task on the quantum annealer only needs to run once for each master problem. Hence, the number of qubits is large enough for the master problem. The number of iterations in the Benders decomposition is estimated to be equal to the case Benders-Gurobi. Furthermore, we see significant improvement potential in our programming of the Benders decomposition, e.g., by using a different programming language than Python. Python frameworks have often the potential to run much faster by different adaptations, e.g., parallel computing or changing the programming language, e.g., C, C++ or Julia (Prechelt, 2000; Fourment and Gillings, 2008; Fua and Lis, 2020). For the best case, an improvement potential of 90 % is estimated. The reduction in computational time we only applied to the python time, so the time spent outside of the optimization solvers Gurobi, D-Wave and Simulated Annealing. Based on these best-case assumptions, the estimated computation times are given in Figure 4. The best-case computational times show that the Benders decomposition using the qunatum annealer and the Benders decomposition using Gurobi have equal computational times. The Benders decompostion with simulated annealing has the highest computational times. Still, in our case study, the Benders decomposition using quantum annealers cannot outperform Gurobi. Furthermore, Gurobi becomes advantageous with increasing problem instance, i.e., number typical time series. In consequence, we do not expect that the Benders decomposition approach using quantum annealers will outperform Gurobi with today's and near future quantum computers. As a noteworthy limitation, our study could only analyze a limited number of instances due to the limited availability and expensive cost of quantum computers today. However, this limitation reflects the current state of quantum computing, where resources are scarce due to their cost. Furthermore, it is important to understand that with today's small and error-prone quantum annealers, achieving convergence becomes increasingly difficult with increasing problem size (Ding et al., 2021). This finding suggests that quantum-annealing-based approaches showing limited performance for small problem instances cannot be expected to perform better for larger instances today. Due to the strong extrapolations made, our conclusions need to be reinvestigated on much larger problem instances, if much larger and fault-tolerant quantum annealers are available.

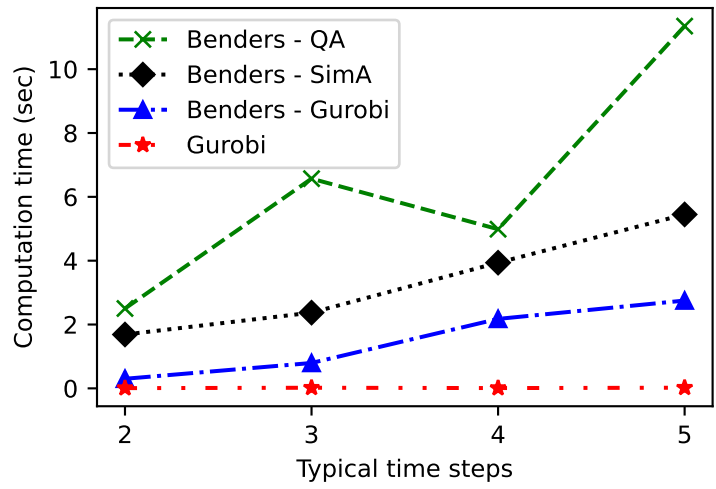

**Fig. 2.** Computation time of the compared approaches to solve the multi-energy system design problem for different numbers of typical time steps.

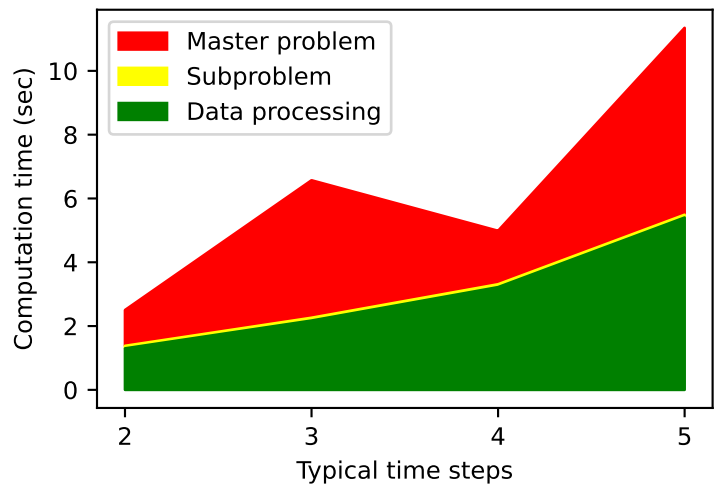

**Fig. 3.** Breakdown of computation time for typical time steps considered in the Benders decomposition using the quantum-computing approach Benders-QA. We separate the computation time in effort required for the master problem, subproblem and time spent on data processing.

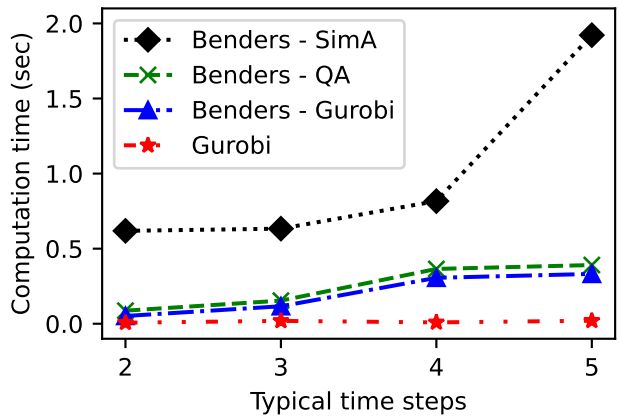

**Fig. 4.** Estimated optimistic computational time of the compared approaches to solve energy system design problems with reduced Python time and improved quantum annealers (details in text).

## 4 Conclusions

A Benders decomposition approach is proposed to integrate quantum computers and classical computers to solve MILP problems for the design of multi-energy systems. The original MILP problem is decomposed such that the quantum computer solves the master problem, where all integer decisions of the original MILP are optimized. The subproblem is solved on a classical computer, where all continuous decisions are taken.

The approach is implemented using a quantum annealer from D-Wave and Gurobi. A multi-energy system is designed for different numbers of typical time steps to analyze the effect of increasing problem size. The proposed hybrid quantum approach is shown to yield suitable designs according to the specifications. However, we observe that hybrid quantum Benders decomposition is clearly outperformed by the classical computer using Gurobi.

Even when taking a best-case assumption on the python time and on the quantum annealer performance, classical optimization approaches are expected to outperform the proposed hybrid quantum approach. If we extrapolate our findings, we do not expect that, in the field of energy system design optimization, the Benders decomposition approach using quantum annealers will beat classical computers in the near future. Still, we could only solve a limited number of instances due to limited availability of the quantum annealer which limits the generality of the results. Thus, this finding should be revisited once more resources become available. However, if fault-tolerant and very large quantum computers are available, this conclusion needs to be tested for large problem sizes that are challenging even for classical computers. In this case, quantum computers can be tested with other approaches, e.g., without a decomposition by discretizing the continuous variables and solving the optimization problem entirely on the quantum computer.

# CRediT authorship contribution statement

**Ludger Leenders:** Conceptualization, Methodology, Investigation, Visualization, Project administration, Funding acquisition, Writing - Original Draft. **Martin Sollich:** Methodology, Investigation, Visualization, Software, Data Curation, Funding acquisition, Writing - Original Draft. **Christiane Reinert:** Investigation, Writing - Review & Editing. **André Bardow:** Conceptualization, Writing - Review & Editing, Supervision, Resources, Funding acquisition.

# Declaration of Competing Interest

The authors declare the following financial interests/personal relationships which may be considered as potential competing interests: Martin Sollich reports equipment, drugs, or supplies was provided by Amazon Web Services Inc. Martin Sollich reports travel was provided by IDEA League. Ludger Leenders reports financial support was provided by Swiss Federal Office of Energy. André Bardow reports financial support was provided by Swiss Federal Office of Energy.

# Acknowledgments

The research published in this publication was carried out with the support of the Swiss Federal Office of Energy as part of the SWEET PATHFNDR. The authors bear sole responsibility for the conclusions and the results presented in this publication.
Ludger Leenders and André Bardow acknowledge the funding by the Swiss Federal Office of Energy's SWEET programme as part of the project PATHFNDR.
Martin Sollich acknowledges the support by the IDEA League and the AWS Cloud Credit for Research program.